# Field-Dependent Kinetics, Exact Kinks, and a Moduli-Space Route from Self-Dual Yang–Mills Theory

Ji Wang
Piezoelectric Device Laboratory, School of Mechanics and Engineering Science,
Ningbo University, 818 Fenghua Road, Ningbo, 315211 Zhejiang, China
E-mail: wangji@nbu.edu.cn

November 23, 2026

## Abstract

We study scalar-field models in one time and one space dimension whose temporal and spatial kinetic coefficients depend on the field itself, motivated by the role of self-dual Yang–Mills (SDYM) theory as a generating framework for lower-dimensional integrable equations. Starting from the generalized Lagrangian, we derive its Euler–Lagrange equation, the traveling-wave reduction, and the first integral, and we solve the inverse compatibility problem: given a kinetic structure and a prescribed hyperbolic-tangent kink, which potential makes that profile an exact solution? Placing the quadratic deformation in the spatial kinetic function, with a canonical temporal sector, yields a sextic reconstructed potential that deforms static configurations directly. We obtain: (i) a closed-form kink energy, with a first-order (Bogomolny-type) formulation, that is independent of the odd deformation parameter although that parameter shapes the energy density asymmetrically; (ii) vacuum curvature relations that tie the fluctuation spectrum to the kinetic function evaluated at the vacua; (iii) two propositions on speed families — a fixed potential supports a prescribed profile at a unique propagation speed unless the temporal kinetic function vanishes identically, and a Lorentz-type family of rescaled kinks exists exactly when the two kinetic functions are proportional — together with a lemma showing that data at a single speed probe only one combination of the two functions; (iv) a unified admissibility analysis in which positivity of the spatial kinetic function simultaneously controls hyperbolicity, the potential bound along the kink, and the vacuum masses; and (v) a self-adjoint fluctuation operator with a closed-form potential whose asymptotes are exactly the vacuum masses, an exact translation zero mode shared by the whole reconstructed family, and numerical spectra with no negative mode anywhere on the admissible parameter region — evidence for spectral stability. Finally, we sharpen the Yang–Mills connection into a precise and honest form: exact SDYM symmetry reductions yield integrable equations, whereas the deformed model is generically non-integrable and cannot be an exact reduction. The correct bridge is the moduli-space route: SDYM reduces dimensionally to the Bogomolny equations, whose BPS monopole and dyon dynamics on moduli space produce precisely field-dependent kinetic terms; any smooth one-dimensional moduli metric expands, to second order, to the quadratic model studied here, and the Taub–NUT metric governing well-separated dyon relative motion provides an explicit realization. The complete identification of the sextic potential within a specific gauge-theory moduli dynamics remains open and is formulated as a concrete research program.

# 1 Introduction

Self-dual Yang–Mills (SDYM) theory in four dimensions occupies a distinguished position in the theory of integrable systems. Under suitable symmetry reductions and gauge choices, the self-duality equations yield a wide class of lower-dimensional nonlinear equations, including several well-known soliton equations [1–4]. This relationship provides a geometric organizing principle for integrable wave motion and has motivated many searches for exact soliton solutions in dimensions below four.

A separate, well-developed line of research concerns scalar field theories with non-canonical kinetic terms, in which the coefficients multiplying the derivatives of the field depend on the field itself. Such structures arise in k-essence cosmology [5], in tachyon effective actions [6], and in the K-field kink literature [7,8]. Field-dependent kinetic functions modify the local propagation speed, the energy density, and the existence and stability properties of kink solutions [7,9].

The question addressed in this paper is a *compatibility problem*: given a field-dependent kinetic structure and a prescribed traveling profile, what scalar potential makes that profile an exact solution? This inverse approach avoids the common but unverified assumption that a canonical soliton survives a deformation of the kinetic sector unchanged. It is the natural complement to the forward constructions of [7,8], where the kinetic structure and potential are fixed first and the solutions are sought afterwards.

A central methodological point, which distinguishes the present work from earlier drafts of this program, concerns *where* the kinetic deformation is placed. In a model with temporal kinetic function $K(\phi)$ and spatial kinetic function $G(\phi)$, a static kink probes only the spatial sector: at zero velocity the temporal function drops out of the quadrature entirely. Placing the deformation in the temporal function therefore makes the static sector canonical and pushes all deformation effects into moving configurations, which is physically opaque — and, as we show in Section 4, a single traveling wave cannot even distinguish the two functions. We therefore take the deformation in the *spatial* function $G(\phi) = 1 + \alpha\phi + \beta\phi^2$ as the primary construction: static kinks are then genuinely deformed, the energy and the fluctuation operator are non-trivial and exactly computable, and every result admits a clean closed form. The temporal deformation is treated as the complementary moving case, and its relation to the primary one is made precise by two propositions.

The main results of this paper are:

1. the general traveling-wave first integral for the two-function model, and the identification of the single combination $G - v^2 K$ that a solution at one speed determines (Section 2, Lemma 1);

2. the explicit reconstructed sextic potential supporting a prescribed tanh kink in the quadratically deformed spatial model, with the canonical quartic double well recovered as the undeformed limit (Section 3);

3. a first-order (Bogomolny-type) formulation and the closed-form kink energy $E = E_0\,(1 + \beta A^2/5)$, which is *independent* of the odd deformation parameter $\alpha$, although $\alpha$ shapes the energy density asymmetrically (Section 3);

4. the vacuum curvature relations $m_\pm^2 = 4G(\pm A)/\Delta^2$, linking the fluctuation spectrum directly to the kinetic function at the vacua (Section 3);

5. two propositions on speed families: uniqueness of the propagation speed for a fixed profile and potential, and the characterization $K = c\,G$ of theories admitting a Lorentz-type rescaled family (Section 4);

6. a unified admissibility analysis in which the single condition $G > 0$ on the field interval controls hyperbolicity, positivity of the spatial energy, the bound $V \geq V_0$ along the kink, and the vacuum masses, together with the exact global condition $\beta > 0,\ \alpha^2 \leq 4\beta$ (Section 5);

7. a self-adjoint fluctuation operator about the deformed kink with a closed-form potential, exact translation zero mode, asymptotes equal to the vacuum masses, and a numerical spectral analysis showing no negative mode anywhere on the admissible region — evidence, at the level available to a finite-difference computation, for spectral stability (Section 6);

8. a precise and honest formulation of the Yang–Mills connection: exact SDYM reductions produce *integrable* equations, the deformed model is generically non-integrable and hence cannot be an exact reduction, and the correct bridge is the moduli-space dynamics of BPS monopoles and dyons descended from SDYM by dimensional reduction, of which the quadratic model is the universal second-order truncation (Section 7).

The paper is deliberately explicit about the boundary between what is established and what remains open. The scalar-field results (items 1–7) are exact and verified both analytically and numerically. The Yang–Mills program (item 8) is formulated as a concrete route — through moduli-space metrics, with the Taub–NUT metric as an explicit realization — rather than claimed as a completed derivation.

The structure of the paper follows the order of the results above. Section 2 introduces the model and the canonical reference case. Section 3 develops the inverse construction. Section 4 analyzes propagation and speed families. Section 5 states the admissibility conditions. Section 6 treats fluctuations and stability. Section 7 develops the Yang–Mills connection. Section 8 contains the discussion and conclusions. Appendix A gives the direct verification of the inverse potential, Appendix B the exact positivity test, and Appendix C the derivation of the closed-form fluctuation potential.

# 2 Generalized scalar model

## 2.1 Lagrangian and field equation

Consider a real scalar field $\phi(t, x)$ in $1 + 1$ dimensions with Lagrangian density

$$\mathcal{L} = 1/2\, K(\phi)\, \phi_t^2 - 1/2\, G(\phi)\, \phi_x^2 - V(\phi), \tag{1}$$

where $K(\phi) > 0$ and $G(\phi) > 0$ are the temporal and spatial kinetic functions and $V(\phi)$ is the scalar potential. The canonical model is recovered for $K = G = 1$. Throughout, primes on functions of $\phi$ denote differentiation with respect to $\phi$.

Variation of Eq. (1) gives the Euler–Lagrange equation

$$K(\phi)\, \phi_{tt} + 1/2\, K'(\phi)\, \phi_t^2 - G(\phi)\, \phi_{xx} - 1/2\, G'(\phi)\, \phi_x^2 + V'(\phi) = 0. \tag{2}$$

The principal part is hyperbolic wherever $K$ and $G$ are positive and regular; small perturbations about a constant background $\phi_0$ propagate with squared speed

$$c^2(\phi_0) = \frac{G(\phi_0)}{K(\phi_0)}, \tag{3}$$

the local characteristic speed of the model.

## 2.2 Traveling-wave reduction and first integral

Let $\phi(t, x) = \phi(\xi)$ with $\xi = x - vt$ be a traveling wave of speed $v$, so that $\phi_t = -v\phi'$, $\phi_{tt} = v^2\phi''$, and $\phi_{xx} = \phi''$, where now primes denote differentiation with respect to $\xi$. Equation (2) becomes

$$\left(v^2 K(\phi) - G(\phi)\right) \phi'' + 1/2 \left(v^2 K'(\phi) - G'(\phi)\right) \phi'^2 + V'(\phi) = 0. \tag{4}$$

Multiplication by $\phi'$ produces a total derivative, because $K$ and $G$ depend on $\phi$ only. One integration gives the quadrature

$$1/2 \left(G(\phi) - v^2 K(\phi)\right) \phi'^2 = V(\phi) - V_0, \tag{5}$$

with integration constant $V_0$. For a kink connecting two degenerate vacua $\phi \to \phi_\pm$ as $\xi \to \pm\infty$, both sides must vanish at the ends, so $V_0 = V(\phi_\pm)$ and $\phi_\pm$ are zeros of $V - V_0$. Equation (5) is the central identity of this paper: the profile, the kinetic functions, and the potential are jointly constrained by it, and it is the engine of the inverse construction below.

A first structural observation, which will be sharpened into Lemma 1, is that (5) contains $K$ and $G$ only through the single combination

$$H_v(\phi) \equiv G(\phi) - v^2 K(\phi). \tag{6}$$

A traveling solution observed at one speed constrains only $H_v$, not $K$ and $G$ separately.

### 2.3 Canonical reference model

For orientation we recall the canonical case $K = G = 1$ with the quartic double well

$$V(\phi) = \frac{\lambda}{4}(\phi^2 - A^2)^2, \qquad V_0 = 0, \tag{7}$$

whose kink and antikink are

$$\phi_{\mathrm{k}}(x) = A\tanh\frac{x}{\Delta}, \qquad \Delta^2 = \frac{2}{\lambda A^2}, \tag{8}$$

with energy $E_0 = 4A^2/(3\Delta) = (2\sqrt{2}/3)\,A^3\sqrt{\lambda}$, and a topological charge which, normalized as $Q = [\phi(\infty) - \phi(-\infty)]/(2A)$, takes the values $Q = \pm 1$ [9]. Linearizing about the kink gives the Pöschl–Teller fluctuation problem

$$\left[-\frac{\mathrm{d}^2}{\mathrm{d}x^2} + \frac{4}{\Delta^2}\left(1 - 3/2\ \mathrm{sech}^2\frac{x}{\Delta}\right)\right]\psi = \omega^2\psi, \tag{9}$$

with discrete frequencies $\omega^2\Delta^2 = 0$ (translation) and $\omega^2\Delta^2 = 3$ (shape mode), and a continuum starting at $\omega^2\Delta^2 = 4$, the squared vacuum mass $m^2 = V''(\pm A) = 4/\Delta^2$. These values serve as benchmarks for the deformed analysis of Section 6.

**Lemma 1 (Single-speed identifiability).**

Fix a speed $v$ and suppose the pair $(K_1, G_1)$ produces, via Eq. (5), the same profile $\phi(\xi)$ and the same potential $V$ as the pair $(K_2, G_2)$ at that speed. Then

$$(K_1 - K_2)(\phi) = \frac{1}{v^2}(G_1 - G_2)(\phi) \qquad \text{on the field range traversed.} \tag{10}$$

In particular, all pairs with the same $G - v^2 K$ are indistinguishable to data at a single speed. For example, the temporal deformation $(K, G) = (1 + \alpha\phi + \beta\phi^2,\ 1)$ at speed $v$ produces exactly the same profile and potential as the spatial deformation $(K, G) = (1,\ 1 - v^2\alpha\phi - v^2\beta\phi^2)$ at the same speed.

**Proof.** Both pairs satisfy Eq. (5) with the same $\phi'(\xi)$, which is nonzero in the interior of a kink, and the same $V - V_0$. Subtracting the two quadratures gives $(v^2K_1 - G_1) - (v^2K_2 - G_2) = 0$ on the traversed field range. ■

Lemma 1 has two consequences for this work. First, the choice of which function to deform is a *choice of parametrization* for single-speed data, and the physically transparent choice for *static* kinks is the spatial one: at $v = 0$ the temporal function drops out of Eq. (5) altogether, so a deformation placed only in $K$ leaves the entire static sector canonical. Second, disentangling $K$ from $G$ requires either multi-speed data (Propositions 1 and 2) or an independent principle — in Section 7 the independent principle is supplied by Yang–Mills moduli-space dynamics, which fixes the effective metric.

# 3 Inverse construction with spatial deformation

## 3.1 Kinetic function and prescribed profile

We now specialize to the primary model of this paper: canonical temporal sector $K = 1$ and quadratic spatial deformation

$$G(\phi) = 1 + \alpha\phi + \beta\phi^2, \qquad \alpha \in \mathbb{R}, \quad \beta \geq 0 \text{ in the admissible region}, \tag{11}$$

and prescribe the hyperbolic-tangent kink

$$\phi_{\mathrm{k}}(x) \;=\; A\tanh\frac{x}{\Delta}\,, \qquad \phi_{\mathrm{k}}{}' \;=\; \frac{A}{\Delta}\left(1-\frac{\phi_{\mathrm{k}}^2}{A^2}\right). \tag{12}$$

Here $A > 0$ is the vacuum offset and $\Delta > 0$ the kink width; both are free design parameters at this stage. The question is: for which potential $V$ is Eq. (12) an exact *static* solution?

### 3.2 Reconstructed potential

For static configurations the quadrature in Eq. (5) reduces to $1/2\, G(\phi_{\mathrm{k}})\phi_{\mathrm{k}}{}'^2 = V(\phi_{\mathrm{k}}) - V_0$. Substituting Eq. (12) and solving for $V$ gives the compact inverse-construction result

$$V(\phi) - V_0 \;=\; \frac{A^2}{2\Delta^2}\,(1+\alpha\phi+\beta\phi^2)\left(1-\frac{\phi^2}{A^2}\right)^2 \tag{13}$$

which is exact for the specified profile, kinetic function, and static character. It is convenient to set $V_0 = 0$ (the vacua then have zero energy). Expanded,

$$V(\phi) \;=\; v_0 + v_1\phi + v_2\phi^2 + v_3\phi^3 + v_4\phi^4 + v_5\phi^5 + v_6\phi^6\,, \tag{14}$$

with

$$v_0 = \frac{A^2}{2\Delta^2}\,, \qquad v_1 = \frac{\alpha A^2}{2\Delta^2}\,, \qquad v_2 = \frac{A^2}{2\Delta^2}\Big(\beta$$
$$v_4 = \frac{A^2}{2\Delta^2}\left(\frac{1}{A^4}-\frac{2\beta}{A^2}\right), \qquad v_5 = \frac{\alpha}{2A^2\Delta^2}\,, \qquad v_6 = \frac{\beta}{2A^2\Delta^2}\,. \tag{15}$$

The potential is generically of degree six: the quadratic kinetic deformation raises the quartic canonical well by two units. In general, a polynomial kinetic deformation of degree $n$ produces a reconstructed potential of degree $n+4$.

### 3.3 Canonical limit and vacuum structure

For $\alpha, \beta \to 0$, Eqs. (13)–(15) reduce to

$$V(\phi) \;\to\; \frac{A^2}{2\Delta^2} - \frac{\phi^2}{\Delta^2} + \frac{\phi^4}{2A^2\Delta^2} \;=\; \frac{\lambda}{4}(\phi^2 - A^2)^2\,, \qquad \lambda = \frac{2}{A^2\Delta^2}, \tag{16}$$

which is precisely the quartic double well with kink width in Eq. (8). The inverse construction therefore contains the standard $\phi^4$ kink as its undeformed limit, and the deformation parameters $(\alpha, \beta)$ deform *both* the potential shape and the kink energetics continuously away from it.

The vacuum structure follows in closed form. Since the factor $(1 - \phi^2/A^2)^2$ has double zeros at $\phi = \pm A$, we have $V(\pm A) = V_0$ and $V'(\pm A) = 0$ for all $(\alpha, \beta)$: the endpoints are always stationary points. Their curvature is

$$m_\pm^2 \;\equiv\; V''(\pm A) \;=\; \frac{4}{\Delta^2}\, G(\pm A) \;=\; \frac{4}{\Delta^2}\,(1 \pm \alpha A + \beta A^2)\,. \tag{17}$$

The squared vacuum masses are directly proportional to the kinetic function evaluated at the vacua. This is a structural result of the inverse construction: the kinetic deformation is imprinted on the fluctuation spectrum through the asymptotic states, a point to which Section 6 returns. The vacua are local minima if and only if $G(\pm A) > 0$, which is one facet of the unified admissibility condition of Section 5.

### 3.4 First-order formulation and closed-form kink energy

The static energy of a configuration interpolating between the vacua is

$$E \;=\; \int_{-\infty}^{\infty} [1/2\; G(\phi)\, \phi_x^2 + V(\phi) - V_0]\, \mathrm{d}x\,. \tag{18}$$

For $G > 0$ along the path it can be completed into a square. Defining the superpotential gradient $W'(\phi) = \sqrt{2G(\phi)(V(\phi) - V_0)}$ (with $W$ fixed up to an additive constant), the energy becomes

$$E \;=\; \frac{1}{2}\int G(\phi)\big(\phi_x \mp W'(\phi)/G(\phi)\big)^2 \mathrm{d}x \;+\; |W(\phi_+) - W(\phi_-)|\,, \tag{19}$$

so that $E \geq |W(A) - W(-A)|$, with equality precisely on solutions of the first-order system $\phi_x = \pm\sqrt{2(V - V_0)/G}$ — which is exactly the quadrature of Eq. (5) at $v = 0$. The reconstructed kink is therefore a saturating (first-order) solution, in the same sense in which the canonical $\phi^4$ kink saturates the Bogomolny bound, although — unlike the topological canonical case — the value $W(A) - W(-A)$ itself depends on the deformation and is not fixed by the boundary data alone.

For the potential of Eq. (13) the superpotential evaluates in elementary terms,

$$W(\phi) \;=\; \frac{A}{\Delta}\left[\phi + \frac{\alpha}{2}\phi^2 + \left(\frac{\beta}{3} - \frac{1}{3A^2}\right)\phi^3 - \frac{\alpha}{4A^2}\phi^4 - \frac{\beta}{5A^2}\phi^5\right], \tag{20}$$

and the kink energy follows in closed form:

$$E \;=\; W(A) - W(-A) \;=\; \frac{4A^2}{3\Delta}\left(1 + \frac{\beta A^2}{5}\right) \;=\; E_0\left(1 + \frac{\beta A^2}{5}\right). \tag{21}$$

Two features deserve emphasis. First, the energy is *independent of* $\alpha$: the odd (symmetry-breaking) deformation reshapes the energy density,

$$\begin{aligned}\rho(x) \;&=\; G(\phi_{\mathrm{k}})\,{\phi_{\mathrm{k}}}'^2 \\ &=\; \frac{A^2}{\Delta^2}(1 + \alpha A \tanh x/\Delta + \beta A^2 \tanh^2 x/\Delta)\operatorname{sech}^4\frac{x}{\Delta},\end{aligned} \tag{22}$$

displacing it towards one vacuum (Figure 2), but its integral is blind to $\alpha$ because the odd contribution integrates to zero over the symmetric field interval. Second, the energy grows linearly in $\beta A^2$: the even deformation stiffens the kink. Both statements are exact.

More generally, for any static solution of the quadrature the energy above the vacuum reduces to a single term,

$$E - V_0 L \;=\; \int G(\phi)\,\phi_x^2\,\mathrm{d}x \;=\; \int_{\phi_-}^{\phi_+} \sqrt{2G(\phi)(V(\phi) - V_0)}\;\mathrm{d}\phi\,, \tag{23}$$

with $L$ the box size; this identity is the workhorse behind Eq. (21) and will also fix the energy of the temporal model in Section 4.

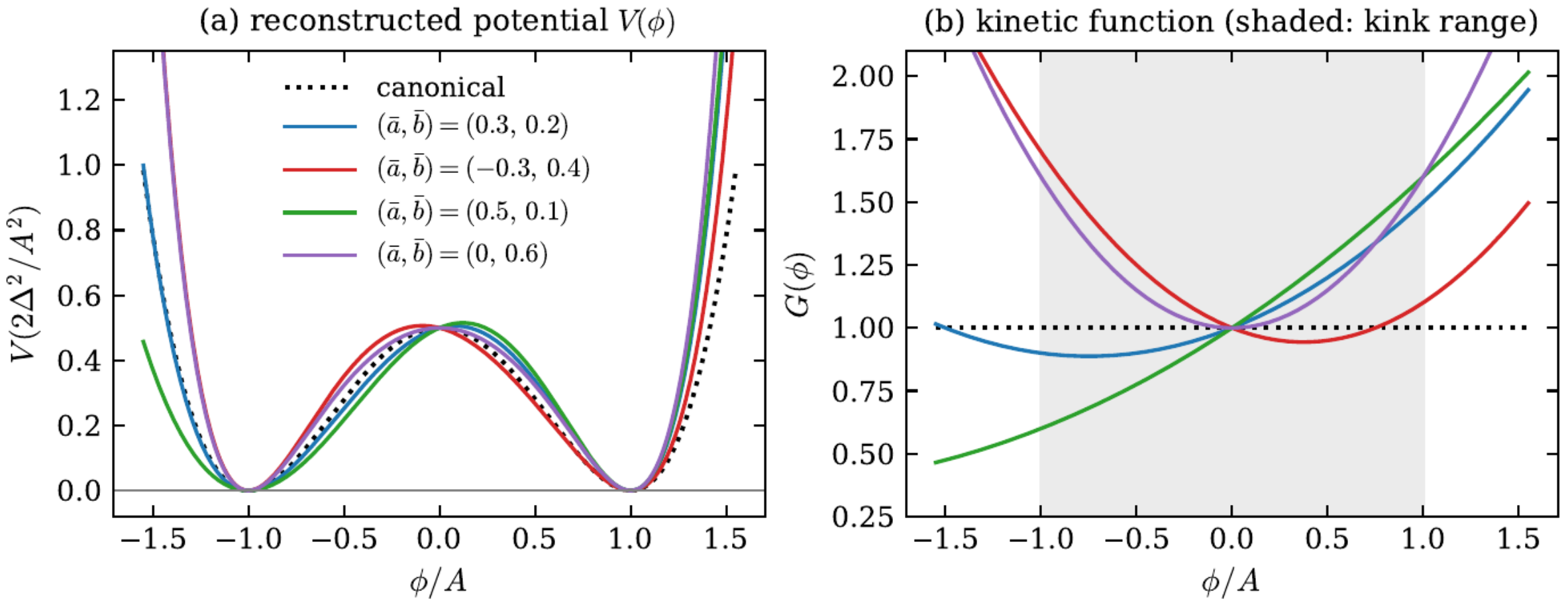

Figure 1: Reconstructed sextic potential in Eq. (13) (a) and kinetic function in Eq. (11) (b) for representative parameter sets, in dimensionless combinations $\bar{a} = \alpha A$, $\bar{b} = \beta A^2$. The dotted black curve is the canonical quartic limit. In (b) the shaded band marks the kink field range $|\phi| \le A$; the kinetic function stays positive there for all shown sets. The odd parameter $\alpha$ tilts both curves asymmetrically.

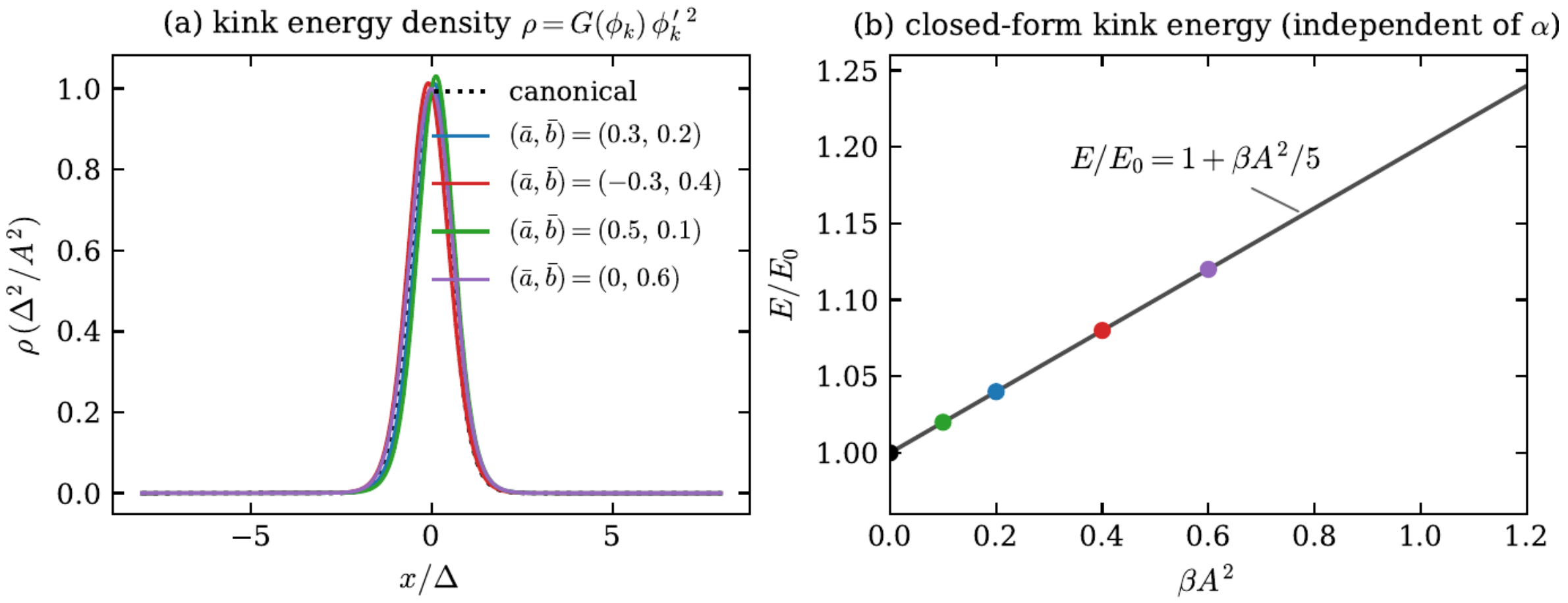

Figure 2: (a) Kink energy density in Eq. (22) for the same parameter sets as Figure 1: $\alpha \neq 0$ displaces the density asymmetrically while leaving the total energy unchanged. (b) The closed-form energy in Eq. (21), $E/E_0 = 1 + \beta A^2/5$, with the sample points marked; the energy depends only on the even deformation parameter.

# 4 Propagation, speed families, and identifiability

## 4.1 Uniqueness of the propagation speed

Lemma 1 shows that single-speed data cannot separate $K$ from $G$. The converse question is sharper: given a *fixed* theory $(K, G, V)$, can a prescribed profile travel at more than one speed?

**Proposition 1 (Unique speed).**

Let the profile $\phi(\xi)$ with $\phi'(\xi) > 0$ on the interior of its field range, and the potential $V$, satisfy the quadrature in Eq. (5) at two distinct speeds $v_1 \neq v_2$ in the same theory. Then

$$(v_1^2 - v_2^2)\, K(\phi)\, \phi'(\xi)^2 \;=\; 0 \qquad \text{for all } \xi\,, \tag{24}$$

and hence $K \equiv 0$ on the traversed field range. In particular, in any non-degenerate model ($K$ nowhere vanishing), a given profile in a given potential propagates at a unique speed.

**Proof.** Write the quadrature at the two speeds and subtract: $1/2\,(G - v_1^2 K)\phi'^2 = V - V_0 = 1/2\,(G - v_2^2 K)\phi'^2$, whence $(v_1^2 - v_2^2)K\phi'^2 = 0$. Since $\phi'^2 > 0$ in the interior, $K$ vanishes identically on the range. ■

## 4.2 When does a speed family exist?

The canonical kink evades Proposition 1 by changing its width with speed: $\phi \mapsto \phi_\gamma(\gamma x - \gamma v t)$ is the Lorentz-dilated family. The following proposition characterizes exactly when such a rescaled family exists in the general model.

**Proposition 2 (Speed-family characterization).**

Fix a theory $(K, G, V)$, and suppose the potential supports the tanh family

$$\phi_v(\xi) \;=\; A \tanh\frac{\xi}{\Delta_v}\,, \qquad \xi = x - vt\,, \tag{25}$$

for all speeds $v$ in a nonempty interval, with $v$-dependent widths $\Delta_v$. Then $K = c\,G$ for a constant $c \geq 0$, and

$$\Delta_v^2 \;=\; \Delta_0^2\,(1 - c\,v^2)\,, \qquad v^2 < 1/c\,, \tag{26}$$

with $\Delta_0$ the static width. Conversely, if $K = c\,G$ and $V$ is the reconstructed potential in Eq. (13) with width $\Delta_0$, then Eqs. (25)– (26) solve the quadrature for every $v^2 < 1/c$. The case $c = 1$ is the Lorentz-invariant model $K = G$.

**Proof.** Substituting Eq. (25) into Eq. (5) gives $1/2\,(G - v^2K)\,(A^2/\Delta_v^2)(1 - \phi^2/A^2)^2 = V - V_0$, so the ratio $(G - v^2K)/\Delta_v^2$ equals a fixed function $f(\phi)$ independent of $v$. Equating the expressions at two speeds $v_1, v_2$ and rearranging,

$$G(\phi)\,(\Delta_2^2 - \Delta_1^2) \;=\; K(\phi)\,(\Delta_2^2 v_1^2 - \Delta_1^2 v_2^2)\,, \tag{27}$$

so $K = c\,G$ with $c = (\Delta_2^2 - \Delta_1^2)/(\Delta_2^2 v_1^2 - \Delta_1^2 v_2^2)$ constant on the field range. Inserting $K = cG$ back into the quadrature gives $(1 - cv^2)G/\Delta_v^2 = f(\phi)$; since the reconstructed potential has $f = G/\Delta_0^2$, the widths obey Eq. (26). The converse follows by direct substitution: with $K = cG$ and $V$ given by Eq. (13), $1/2\,(1 - cv^2)G\,(A^2/\Delta_v^2)(1 - \phi^2/A^2)^2 = 1/2\,G(A^2/\Delta_0^2)(1 - \phi^2/A^2)^2 = V - V_0$ holds precisely when Eq. (26) holds. ■

**Corollary 1.**

In the primary model of Section 3 ($K = 1$, $G$ deformed and non-constant), the reconstructed tanh kink is *strictly static*: no rescaled moving family exists.

This is the precise sense in which the spatial deformation breaks Lorentz invariance: the kink is pinned to the lab frame by the field-dependent spatial stiffness. Moving solutions of the same theory exist in principle, but they are not tanh profiles and must be obtained numerically.

### 4.3 The temporal counterpart

It is instructive to run the same inverse construction with the deformation in the temporal sector, $K(\phi) = 1 + \alpha\phi + \beta\phi^2$, $G = 1$, which is the variant studied in the first draft of this program. The quadrature of Eq. (5) at speed $v$ gives

$$V(\phi) - V_0 \;=\; \frac{A^2}{2\Delta^2}(1 - v^2 - v^2\alpha\phi - v^2\beta\phi^2)\left(1 - \frac{\phi^2}{A^2}\right)^2, \tag{28}$$

again a sextic, but one whose deformation coefficients are suppressed by $v^2$ and which depends explicitly on the speed: each $v$ defines a *different* theory. All the conclusions of Propositions 1 and 2 apply verbatim. Two further facts follow from the general identity in Eq. (23), extended to traveling waves.

First, by Lemma 1 at $v = 0$, the *static* sector of the temporal model is completely canonical: $K$ drops out, and the tanh profile of Eq. (28) is not a static solution of that potential. Any statement about the "static energy of the temporal-model kink" is therefore vacuous — the kink simply does not exist at rest. This is the structural reason the spatial deformation was adopted as the primary construction.

Second, the energy that the traveling kink carries above the vacuum is nonetheless exact and remarkably simple. For any solution of the quadrature at speed $v$,

$$\begin{aligned} E - V_0 L \;&=\; \int [1/2\, K\phi_t^2 + 1/2\, G\phi_x^2 + V - V_0]\mathrm{d}x \\ &=\; \int G(\phi)\, \phi_x^2\, \mathrm{d}x\,, \end{aligned} \tag{29}$$

because $\phi_t^2 = v^2\phi'^2$ and the three terms recombine into $G\phi'^2$ on using Eq. (5). For the temporal model ($G = 1$),

$$E \;=\; \int {\phi_\mathrm{k}}'^2\, \mathrm{d}x \;=\; \frac{4A^2}{3\Delta}, \tag{30}$$

independent of $\alpha$, $\beta$, *and* $v$: within the reconstructed family, the traveling kink always carries the canonical energy of its width, the deformation trading kinetic-sector structure for potential-sector structure at fixed laboratory energy.

Table 1 summarizes the two placements of the deformation.

Table 1: Comparison of the two placements of the quadratic kinetic deformation. The spatial placement is adopted in this paper because it deforms the physically transparent static sector; the temporal placement is its moving counterpart, constrained by the same propositions.

| | temporal deformation | spatial deformation |
|---|---|---|
| | $K = 1 + \alpha\phi + \beta\phi^2$, $G = 1$ | $K = 1$, $G = 1 + \alpha\phi + \beta\phi^2$ |
| static sector | fully canonical ($K$ invisible) | deformed — primary construction |
| tanh solution | traveling, at fixed speed $v$ | static |
| reconstructed $V$ | sextic, coefficients $\propto v^2\alpha,\ v^2\beta$, speed-dependent | Sextic in Eq. (13), speed-free |
| energy above vacuum | $4A^2/3\Delta$, independent of $v, \alpha, \beta$ | $E_0(1 + \beta A^2/5)$ (21) |
| Lorentz invariance | broken; no rest frame | broken; kink pinned (Corollary 1) |

# 5 Physical admissibility

## 5.1 One condition, four roles

The reconstructed model is physically admissible only in a subregion of the $(\alpha, \beta)$ plane. Remarkably, four independent requirements reduce to a single condition on $G$ restricted to the kink field interval $[-A, A]$:

1. *Hyperbolicity.* The characteristic speed in Eq. (3) about background values on the kink is real and positive iff $G > 0$ (with $K = 1$).

2. *Positive spatial energy.* The spatial part of the Hamiltonian density, $1/2\, G(\phi)\phi_x^2$, is non-negative on the kink iff $G \geq 0$ on $[-A, A]$.

3. *The bound along the kink.* From the quadrature at $v = 0$, $V(\phi) - V_0 = 1/2\, G(\phi){\phi_{\mathrm{k}}'}^2$, the potential stays at or above the vacuum value along the trajectory iff $G \geq 0$ on $[-A, A]$.

4. *Vacuum minima.* The squared vacuum masses in Eq. (17) are positive iff $G(\pm A) > 0$.

We therefore define:

**Definition 1 (On-trajectory admissibility).**

The deformed model is *admissible on the kink trajectory* if

$$G(\phi) \; > \; 0 \qquad \text{for all } \phi \in [-A, A]\,. \tag{31}$$

For the quadratic $G$ of Eq. (11), condition in Eq. (31) admits an exact elementary test, given in Appendix B. In the dimensionless combinations $\bar{a} = \alpha A$, $\bar{b} = \beta A^2$ the test reads:

$$1 \pm \bar{a} + \bar{b} > 0 \qquad \text{and} \qquad \begin{cases} \bar{a}^2 \le 4\bar{b} \quad \text{or} \quad |\bar{a}| \ge 2\bar{b}\,, & \bar{b} > 0 \ \ (\text{convex } G)\,, \\ (\text{endpoint test only})\,, & \bar{b} \le 0 \ \ (\text{concave or line} \end{cases} \tag{32}$$

The first line says that for convex $G$ the minimum over the interval sits either at the endpoints (tested by $1 \pm \bar{a} + \bar{b}$) or at the interior vertex (tested by $\bar{a}^2 \le 4\bar{b}$, i.e. $G$ has no real roots).

## 5.2 Global admissibility

Condition in Eq. (31) is local to the kink. Two global properties are controlled by the behavior of $G$ on the whole field axis, because $V(\phi) - V_0 = 1/2\, G(\phi) \cdot (A/\Delta)^2 (1 - \phi^2/A^2)^2$ everywhere:

- *Boundedness.* The sextic in Eq. (14) is bounded below iff $\beta > 0$ (leading term $\beta\phi^6/2A^2\Delta^2$), or in the degenerate case $\alpha = \beta = 0$ (quartic). For $\beta = 0 < |\alpha|$ the potential is an unbounded quintic; for $\beta < 0$ it is unbounded from below.

- *Global positivity.* $V \ge V_0$ on all of $\mathbb{R}$ iff $G \ge 0$ on all of $\mathbb{R}$, which for the quadratic holds iff

$$\beta > 0 \qquad \text{and} \qquad \alpha^2 \le 4\beta\,, \tag{33}$$

  together with the canonical point $\alpha = \beta = 0$. When Eq. (33) fails but Eq. (31) holds, the potential dips below $V_0$ somewhere *off* the kink path, and the vacua $\pm A$, while still local minima, are no longer the global minima — the kink then sits above the true ground state and its stability is not guaranteed.

The admissible region is displayed in Figure 3, which also marks the sample parameter sets used throughout the paper and the Taub–NUT truncation point derived in Section 7. All sample sets lie in the globally admissible region $\bar{b} > 0$, $\bar{a}^2 \le 4\bar{b}$.

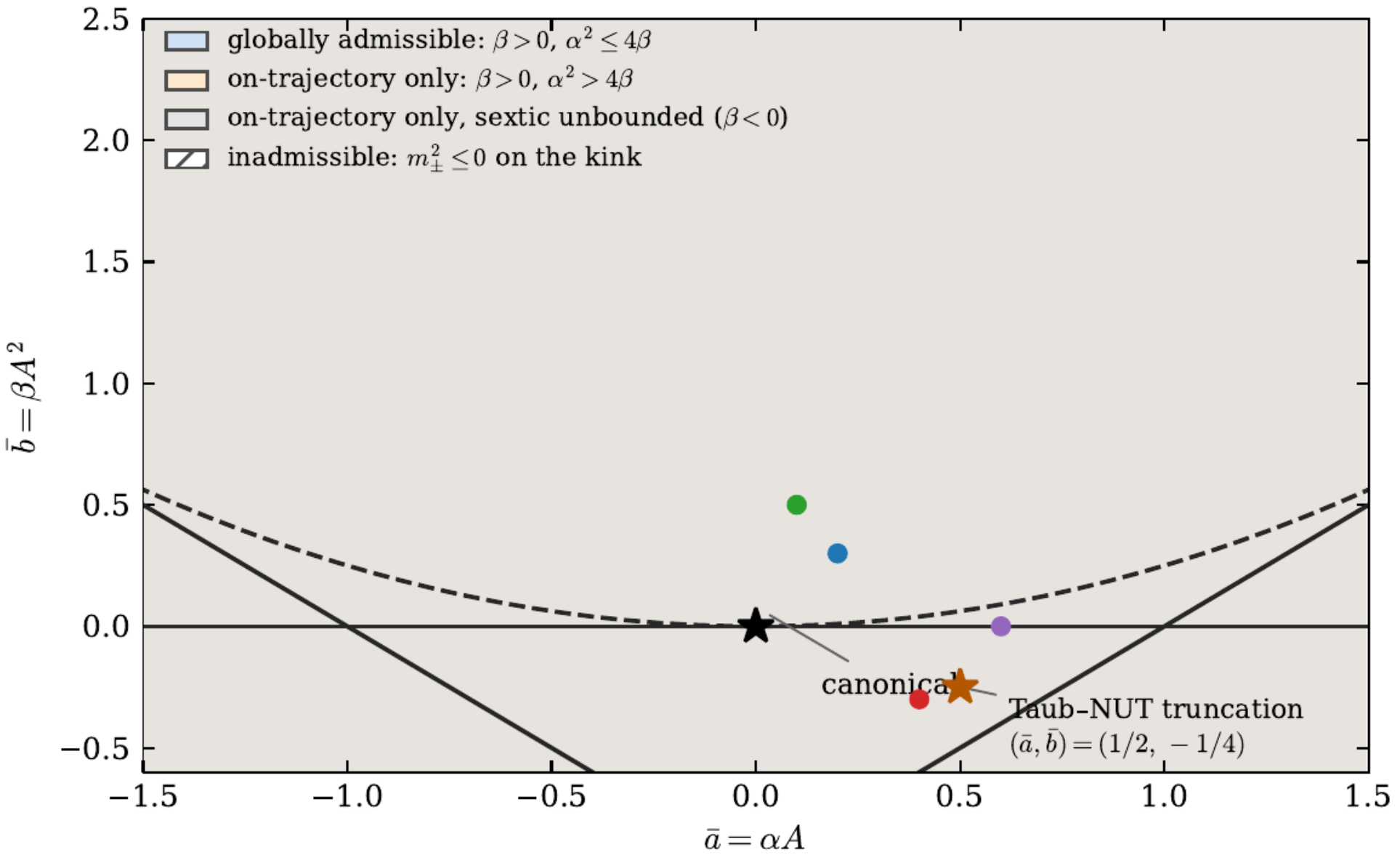

Figure 3: Admissible region of the deformed model in the $(\bar{a},\bar{b}) = (\alpha A, \beta A^2)$ plane. Blue: globally admissible ($\beta > 0$, $\alpha^2 \leq 4\beta$; solid parabola boundary). Orange: on-trajectory admissible only, with $V$ bounded but dipping below $V_0$ off the kink path ($\beta > 0$, $\alpha^2 > 4\beta$). Gray: on-trajectory admissible with unbounded sextic ($\beta < 0$). Hatched: inadmissible (a vacuum mass is non-positive). The star at the origin is the canonical model; colored dots are the sample sets of Figures 1–5; the brown star marks the Taub–NUT moduli-metric truncation of Section 7, which is admissible on the trajectory but not globally — a local truncation need not respect the global conditions of the polynomial model.

### Remark.

The hierarchy of conditions has a practical reading. Numerical and variational work with the deformed kink (Section 6) only requires Eq. (31). Statements about the vacuum structure of the full theory — e.g. that the kink is the lowest nontrivial static sector — require the global condition in Eq. (33). In the moduli-space setting of Section 7, only the on-trajectory condition is guaranteed by the gauge-theory truncation, and this distinction matters.

# 6 Fluctuations and spectral stability

## 6.1 Quadratic action and the fluctuation operator

Write $\phi(t,x) = \phi_{\mathrm{k}}(x) + \eta(t,x)$ and expand the action to second order in $\eta$. The details are given in Appendix C; the result is

$$S^{(2)} = \frac{1}{2}\int \mathrm{d}t \int \mathrm{d}x \,[\eta_t^2 - G(\phi_{\mathrm{k}})\,\eta_x^2 - U(x)\,\eta^2], \tag{34}$$

where the fluctuation potential is

$$U(x) = V''(\phi_{\mathrm{k}}) - 1/2\, G''(\phi_{\mathrm{k}})\,\phi_{\mathrm{k}}'^2 - G'(\phi_{\mathrm{k}})\,\phi_{\mathrm{k}}''. \tag{35}$$

The first term is the familiar curvature of the potential along the kink; the second and third are induced by the field-dependence of the spatial kinetic function and have no canonical counterpart. The fluctuation equation is $\ddot{\eta} = -\mathcal{L}\eta$ with

$$\mathcal{L} = -\frac{\mathrm{d}}{\mathrm{d}x}\, G(\phi_{\mathrm{k}})\,\frac{\mathrm{d}}{\mathrm{d}x} + U(x). \tag{36}$$

Two structural facts make the deformed problem as clean as the canonical one:

- *Self-adjointness.* $\mathcal{L}$ is self-adjoint with respect to the standard flat inner product, $\int f\,\mathcal{L}g\,\mathrm{d}x = \int (\partial_x f)\,G(\phi_{\mathrm{k}})\,(\partial_x g)\,\mathrm{d}x + \int f\,U\,g\,\mathrm{d}x$, for fields decaying at infinity. The frequencies are therefore real, $\omega^2 \in \mathbb{R}$, and stability is equivalent to non-negativity of the spectrum of $\mathcal{L}$. No field redefinition or weighted measure is required — the natural quadratic form of the action is already symmetric.

- *Exact zero mode.* Translation invariance survives the kinetic deformation, and indeed one verifies (Appendix C) that

$$\mathcal{L}\,\phi_{\mathrm{k}}' = 0, \qquad \phi_{\mathrm{k}}'(x) = \frac{A}{\Delta}\mathrm{sech}^2\frac{x}{\Delta}, \tag{37}$$

with a *parameter-independent* shape: all kinks of the reconstructed family, whatever $(\alpha, \beta)$, share the canonical zero mode.

## 6.2 Closed-form fluctuation potential

Substituting Eqs. (11), (12), and (13) into Eq. (35) and simplifying (Appendix C) yields the exact closed form

$$U(x) = \frac{2}{\Delta^2}[(2-3s^2)(1+\alpha A\,t+\beta A^2t^2) - s^2(\alpha A\,t+2\beta A^2t^2)], \tag{38}$$

with $t = \tanh(x/\Delta)$ and $s = \mathrm{sech}(x/\Delta)$. Its properties:

1. *Asymptotics.* As $x \to \pm\infty$ ($s \to 0, t \to \pm 1$),

   $$U(\pm\infty) = \frac{4}{\Delta^2} G(\pm A) = m_\pm^2, \tag{39}$$

   in agreement with Eq. (17): the fluctuation potential approaches exactly the squared vacuum masses, so the continuum thresholds of the deformed problem are the deformed vacuum masses.

2. *Canonical limit.* For $\alpha = \beta = 0$, $U \to (2/\Delta^2)\big(2 - 3\mathrm{sech}^2(x/\Delta)\big)$, the Pöschl–Teller potential of Eq. (9) with discrete spectrum $\omega^2\Delta^2 \in \{0,3\}$ and continuum $[4,\infty)$.

3. *Asymmetry.* For $\alpha \neq 0$ the two asymptotic levels differ and the well is tilted (Figure 4); the spectrum is no longer reflection symmetric, but remains real by self-adjointness.

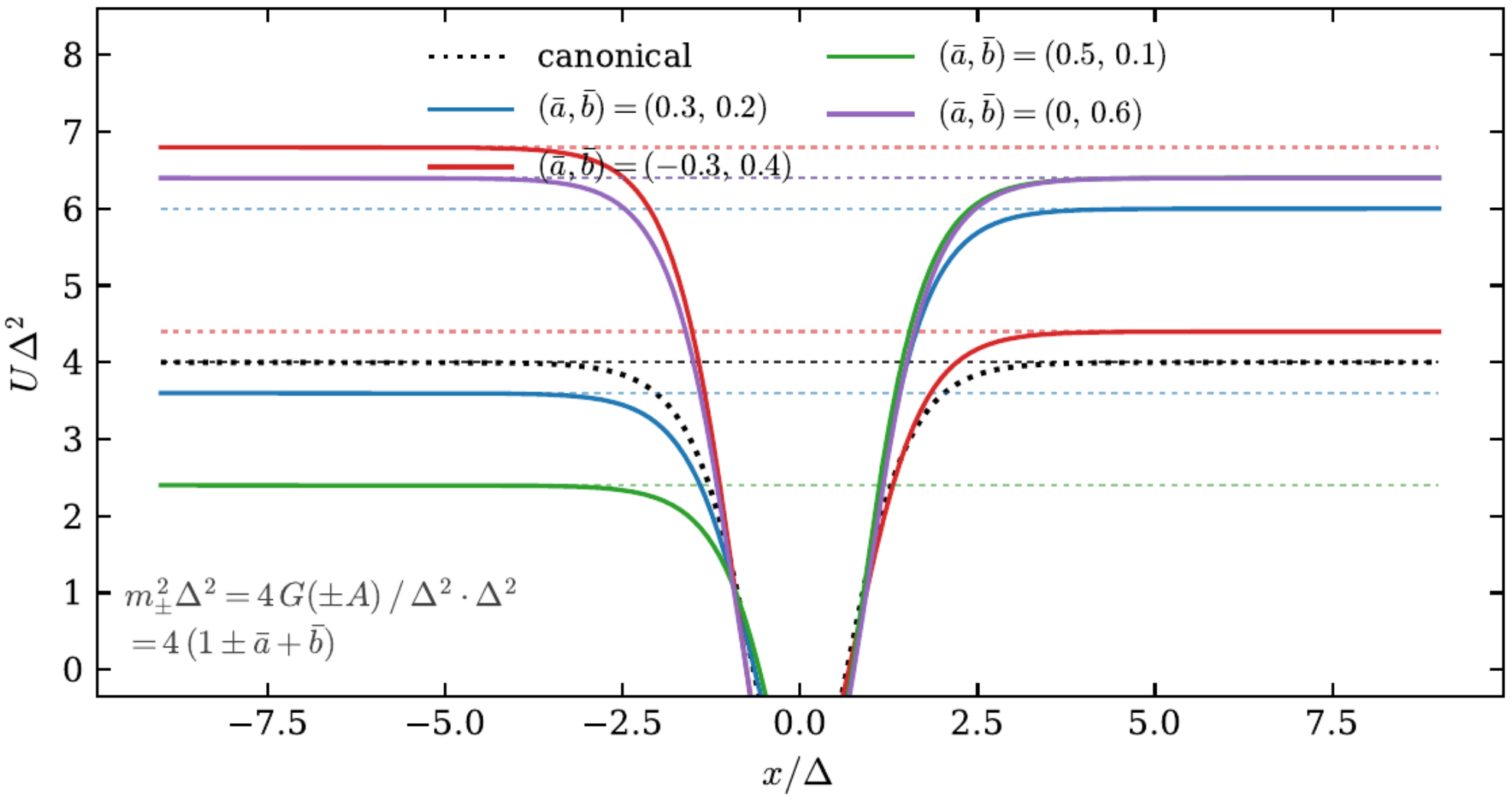

Figure 4: Closed-form fluctuation potential (38) for the sample parameter sets. Dashed horizontal lines mark the asymptotic values $m_\pm^2 \varDelta^2 = 4(1 \pm \bar{a} + \bar{b})$, which are the continuum thresholds. The dotted black curve is the canonical Pöschl–Teller limit.

### 6.3 Numerical spectra

Self-adjointness and the exact zero mode guarantee that $\lambda_0 = 0$ is an eigenvalue, but not that it is the *lowest*. Establishing the absence of negative modes is the substance of spectral stability, and we address it numerically.

The operator in Eq. (36) is discretized by conservative finite differences on $x \in [-25\varDelta, 25\varDelta]$ with $N = 4000$ intervals, evaluating $G$ at midpoints (staggered grid), which preserves the symmetry and the conservative form of $\partial_x(G\,\partial_x)$ ; the resulting tridiagonal matrix is diagonalized exactly. Table 2 lists the three lowest eigenvalues $\lambda_i = \omega_i^2$ for the sample sets, together with the continuum thresholds and energies.

Table 2: Numerical fluctuation spectra of the deformed kink for the sample parameter sets. The canonical row reproduces the exact Pöschl–Teller values $\{0,3\} \cup [4, \infty)$ to five digits. In every case the lowest eigenvalue is the translation zero mode to within the $O(h^2) \sim 3 \times 10^{-5}$ discretization error, the first nonzero eigenvalue is positive, and the higher levels sit at or above the appropriate continuum threshold $m_\pm^2$ (e.g. $\lambda_2 \to m_-^2$ for (0.3,0.2) and → $m_+^2$ for $(-0.3,0.4)$, while $(0,0.6)$ supports a second bound state below the common threshold 6.4).

| $(\bar{a},\bar{b})$ | $\lambda_0\Delta^2$ | $\lambda_1\Delta^2$ | $\lambda_2\Delta^2$ | $m_-^2\Delta^2$ | $m_+^2\Delta^2$ | $E/E_0$ |
|---|---|---|---|---|---|---|
| (0, 0) canonical | $-3.0 \times 10^{-5}$ | 2.9999 | 4.0045 | 4 | 4 | 1 |
| (0.3, 0.2) | $-3.1 \times 10^{-5}$ | 2.9964 | 3.6161 | 3.600 | 6.000 | 1.04 |
| (−0.3, 0.4) | $-3.2 \times 10^{-5}$ | 3.2028 | 4.4179 | 6.800 | 4.400 | 1.08 |
| (0.5, 0.1) | $-3.0 \times 10^{-5}$ | 2.3837 | 2.4161 | 2.400 | 6.400 | 1.02 |
| (0, 0.6) | $-3.3 \times 10^{-5}$ | 3.4651 | 5.9969 | 6.400 | 6.400 | 1.12 |

Three systematic checks were performed. First, the canonical case reproduces the exact spectrum $\{0,3\} \cup [4,\infty)$ (row one of Table 2). Second, the numerical eigenvector of $\lambda_0$ collapses onto the analytic zero mode in Eq. (37) for every parameter set, with maximum deviation $2 \times 10^{-5}$ (Figure 5a). Third, a scan over the admissible region, $\bar{a} \in [-1.2,1.2]$, $\bar{b} \in [0,1.5]$ subject to $1 \pm \bar{a} + \bar{b} > 0$, found *no negative eigenvalue anywhere*; the first nonzero eigenvalue stays positive and degenerates only at the boundary of the admissible region, where a vacuum mass itself vanishes (Figure 5b).

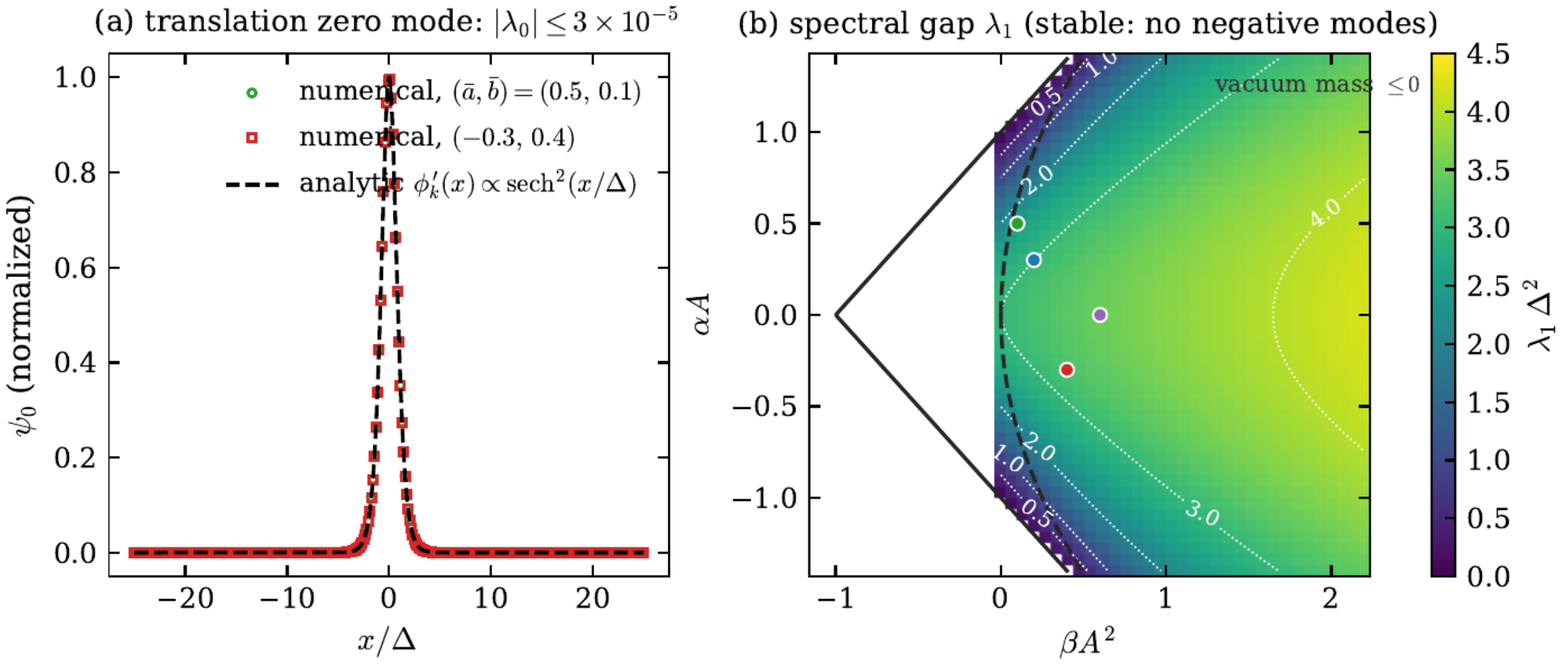

Figure 5: (a) Numerical zero modes (markers) for two parameter sets, collapsing onto the analytic translation mode $\phi_k{}' \propto \mathrm{sech}^2(x/\Delta)$ (dashed line); $|\lambda_0| \leq 3.3 \times 10^{-5}$, of the order of the discretization error. (b) First nonzero eigenvalue $\lambda_1\Delta^2$ over the admissible region (color map; dotted white contours at 0.5,1, … ,4). The gap is positive throughout and closes only where a vacuum mass degenerates (solid line); the dashed parabola is the global-admissibility boundary $\bar{a}^2 = 4\bar{b}$. Dots mark the sample sets.

## 6.4 Status of the stability statement

Within the scope of this paper the stability result is a *numerical* one, and we state it as such: for every admissible parameter set examined, the fluctuation operator in Eq. (36) is positive semi-definite with the expected zero mode, one or more discrete bound states below the deformed continuum thresholds, and no negative mode. The structural facts — self-adjointness, the exact zero mode, the matching asymptotics in Eq. (39) — are exact and parameter-independent, and they make the numerical evidence robust; but a proof of non-negativity, for instance by a factorization or Sturm-comparison argument adapted to the weighted operator $-\partial_x(G\,\partial_x)$, is a well-posed open problem which we do not claim to have solved. We emphasize that no spectral-stability claim beyond this evidence is made anywhere in this paper.

# 7 The Yang–Mills connection

The scalar-model results of Sections 3–6 are self-contained. This section formulates their relation to self-dual Yang–Mills theory as precisely as possible: first by delimiting what exact SDYM reductions can and cannot provide, then by exhibiting the route — moduli-space dynamics of BPS monopoles — through which field-dependent kinetic terms genuinely descend from SDYM, and finally by listing what a complete embedding still requires.

## 7.1 What exact SDYM reductions give — and what they cannot give

The SDYM equations on a four-manifold with coordinates and orientation fixed,

$$F_{\mu\nu} = \pm 1/2 \, \epsilon_{\mu\nu\rho\sigma} F^{\rho\sigma} , \tag{40}$$

are the compatibility condition of a linear system, and under symmetry reductions (translation invariance in chosen directions, coupled with gauge conditions on the connection) they yield large classes of integrable equations in $1+1$ and $2+1$ dimensions: the sine-Gordon, KdV, and nonlinear Schrödinger equations among them [1–4]. This is the precise, established content of the phrase "SDYM as a generating framework."

It is important to be equally precise about the limits of this statement. The equations produced by exact SDYM reductions are *integrable*: they possess Lax pairs, infinite conservation laws, and soliton scattering factorization [2,10]. The deformed model of this paper is, for generic $(\alpha, \beta) \neq 0$, not known to possess any of these structures (and neither does the canonical $\phi^4$ model itself, which is famously non-integrable). It follows that neither the canonical quartic kink nor its sextic deformation can be an exact symmetry reduction of SDYM in the sense of [3,4]. Any suggestion that the model "derives" from SDYM by reduction would be a category error, conflating exact integrable reductions with approximate effective descriptions. We record this explicitly because earlier drafts of this program used looser language.

The correct question is therefore not *reduction* but *effective description*: can the pair (field-dependent kinetic function, deformed potential) arise as the low-energy effective dynamics of degrees of freedom that genuinely live in a Yang–Mills theory descended from SDYM? The answer is yes, along the following chain.

## 7.2 The moduli-space route

The chain has four links, each established in the literature.

*Link 1: dimensional reduction.* Imposing invariance of Eq. (40) along one coordinate reduces four-dimensional self-duality to the Bogomolny equation on $\mathbb{R}^3$ [11],

$$B_i \;=\; \pm D_i\Phi\,, \tag{41}$$

whose static, finite-energy solutions are the BPS monopoles [11–13]. The BPS monopole is thus literally a descendant of SDYM.

*Link 2: moduli-space dynamics.* The low-energy dynamics of BPS solitons is geodesic motion on the moduli space $\mathcal{M}$ of solutions, equipped with the $L^2$ metric [14,15]. For collective coordinates $q^a(t)$,

$$L_{\text{eff}} \;=\; 1/2\; g_{ab}(q)\,\dot{q}^a\dot{q}^b \;-\; \cdots, \tag{42}$$

where the ellipsis denotes terms of higher order in velocities and any perturbations that lift the BPS degeneracy. Restricting to a single collective coordinate $q$, the effective kinetic term is *exactly* of the field-dependent form studied in this paper, with $K = g_{qq}$.

*Link 3: universal second-order truncation.* Any smooth one-dimensional moduli metric $g(q)$ can be expanded about a reference point $q_0$ with $\varphi = q - q_0$:

$$\frac{g(q_0+\varphi)}{g(q_0)} \;=\; 1 + \alpha\varphi + \beta\varphi^2 + O(\varphi^3)\,. \tag{43}$$

The quadratic model in Eq. (11) is therefore the *universal* second-order truncation of one-dimensional moduli-space dynamics descending from SDYM. In this precise sense, the class of kinetic functions studied here is not merely motivated by Yang–Mills theory: it is the generic local form of BPS collective-coordinate dynamics.

*Link 4: an explicit realization.* For a pair of well-separated BPS dyons, the relative part of the moduli-space metric is asymptotically of Taub–NUT type [15–18],

$$\mathrm{d}s^2 = V(r)(\mathrm{d}\psi + \cos\theta\,\mathrm{d}\varphi)^2 + V(r)^{-1}\mathrm{d}r^2 + r^2\mathrm{d}\Omega^2\,, \qquad V(r) = 1 + \frac{2m}{r}, \tag{44}$$

where $m$ is the monopole mass scale. Purely radial relative motion is geodesic motion with effective kinetic function

$$K(r) = \frac{1}{V(r)} = \frac{1}{1 + 2m/r}, \tag{45}$$

a genuinely field-dependent (coordinate-dependent) kinetic coefficient descending entirely from Yang–Mills data. Normalizing by $K(r_0)$ and expanding at $r_0 = 2m$ gives, with $\varphi = r - r_0$,

$$\frac{K(r_0 + \varphi)}{K(r_0)} = 1 + \frac{\varphi}{4m} - \frac{\varphi^2}{16m^2} + O(\varphi^3)\,. \tag{46}$$

Truncating at second order and taking a field range of half-width $A = 2m$ identifies the moduli-derived point

$$(\bar{a}, \bar{b}) = \left(\frac{A}{4m}, -\frac{A^2}{16m^2}\right) = \left(\frac{1}{2}, -\frac{1}{4}\right), \tag{47}$$

which lies in the on-trajectory admissible but *not* globally admissible region of Figure 3 ($\bar{b} < 0$). This is instructive rather than pathological: a local truncation of an exact Yang–Mills metric has no reason to satisfy the global admissibility conditions of the polynomial model it approximates — those conditions pertain to the truncated model, not to the underlying gauge theory. Conversely, the global conditions of Section 5 delimit the region in which the polynomial truncation can be trusted as a model in its own right.

*The role of the potential.* In exact moduli-space dynamics the BPS condition makes the effective potential on $\mathcal{M}$ vanish; a nontrivial $V(\varphi)$ arises when the degeneracy is lifted — by boundaries, external fields, weak symmetry breaking, or higher-order corrections — and it is precisely at this point that the inverse construction of Section 3 becomes operative: given the Yang–Mills-supplied kinetic function of Eq. (43) and a desired switching profile for the collective coordinate, equations (13) and (21) deliver the unique potential that makes that profile exact, along with its energy. The scalar-model machinery and the Yang–Mills supply line thus compose naturally.

**Remark (Field redefinitions).**

The identification of the scalar field with a gauge-theory collective coordinate fixes a normalization. Under a reparametrization $\varphi \mapsto f(\varphi)$ of the collective coordinate, the kinetic function and the potential transform jointly, $K \mapsto K \circ f\,(f')^2$ and $V \mapsto V \circ f$, so the split between "kinetic" and "potential" deformation is parametrization-dependent even though the physics is not. This is the same subtlety, in a sharper form, as Lemma 1: the invariants are the trajectory data, not the parametrization. Any complete embedding must state the normalization explicitly.

## 7.3 What a complete embedding still requires

The above chain establishes, at the level of effective dynamics, that field-dependent kinetic terms are native to Yang–Mills theories descended from SDYM. What it does not establish is the existence, within one specific gauge-theory system, of an effective description whose potential matches the sextic in Eq. (13) and whose kink in Eq. (12) approximates actual collective-coordinate motion. Completing the embedding requires:

1. a definite choice of gauge group, signature, and BPS sector (monopole pair, vortex pair, or calorons);
2. the exact moduli metric along a chosen one-dimensional submanifold, beyond second order, to control the truncation error of Eq. (43);
3. a concrete mechanism lifting the BPS degeneracy, and the effective potential it induces, to be compared with Eq. (13);
4. verification that the reduced action reproduces the reduced equations, with all residual gauge constraints imposed and all field redefinitions tracked;
5. the recovery of our admissibility conditions from positivity of the exact metric, explaining why the on-trajectory condition is inherited naturally (Figure 3) while the global conditions need not be.

These items constitute a well-defined research program. Its output would elevate the moduli-space route from a structural observation to a derivation, and the deformed-kink machinery of Sections 3–6 would then supply the exact soliton content of the resulting effective theory.

# 8 Discussion and conclusions

## 8.1 Relation to the literature

The construction of this paper overlaps, in intent, with three literatures, and its position with respect to each is worth making precise.

*K-field kinks.* Zhong and Liu [7] analyze generalized kinetic terms with the potential fixed a priori and construct the kink solutions that survive; the logic is forward (from model to solution). The present work inverts the logic: the profile is prescribed and the potential is the unknown. The two approaches are complementary, and the fluctuation analysis of Section 6 parallels the stability questions raised there, with the difference that our operator in Eq. (36) is self-adjoint in the flat measure by construction.

*The deformation method.* Bazeia and collaborators [8] generate new defects from known ones by deforming the potential of a canonical-kinetic model. Our inverse construction carries the same "profile-first" philosophy into the kinetic-deformed setting: equation (13) is the explicit solution of that inverse problem for the quadratic kinetic function, and the first-order formulation in Eqs. (19)–(20) mirrors the energy-first structure that makes the deformation method powerful.

*Effective field theories.* k-essence [5] and tachyon effective actions [6] share the field-dependent kinetic structure. Their classical consistency requirements — positive kinetic energy, real propagation speed, subluminality — are the field-theory analogues of the admissibility analysis of Section 5; in the kink sector of our model all of them collapse onto the single condition $G > 0$ on the trajectory, with the global condition in Eq. (33) governing the vacuum structure of the full theory.

## 8.2 Limitations

The honest boundaries of this work are the following. The inverse construction establishes existence and uniqueness of the compatible potential for a chosen profile, kinetic function, and speed; it does not classify solutions for a fixed potential, nor does it address multi-kink configurations, kink–antikink scattering, or radiation. The stability analysis is exact in its structural parts (self-adjointness, zero mode, asymptotics) but numerical in its central claim of non-negativity of the spectrum. The Yang–Mills connection is a supply line for the kinetic sector, demonstrated explicitly through the Taub–NUT example, but the matching of the sextic potential within a specific BPS system remains an open program (Section 7). Finally, the two-flavor structure $K \neq G$ breaks Lorentz invariance, so the usual rest-frame intuition for solitons must be abandoned; we have delimited the consequences in Propositions 1 and 2, but a full Hamiltonian analysis of the moving sectors is beyond the present scope.

### 8.3 Conclusions

We have formulated and solved the inverse compatibility problem for kinks in scalar models with field-dependent kinetic terms. With the quadratic deformation placed in the spatial function $G(\phi) = 1 + \alpha\phi + \beta\phi^2$, the prescribed tanh kink is supported by an explicit sextic potential in Eq. (13) which contains the canonical quartic double well as its undeformed limit. The construction is complete in the strong sense that every quantity of interest is closed-form: the kink energy $E = E_0(1 + \beta A^2/5)$, independent of the odd deformation; the vacuum masses $m_\pm^2 = 4G(\pm A)/\Delta^2$ ; the first-order (Bogomolny-type) formulation in Eq. (19); and the fluctuation potential in Eq. (38) of the self-adjoint operator in Eq. (36), whose asymptotes are exactly the vacuum masses and whose zero mode is exactly the kink derivative for all deformations.

Beyond the exact results, the paper contributes two structural theorems on speed families (unique propagation speed for a fixed profile and potential unless the temporal function vanishes; the characterization $K = c\,G$ of theories admitting rescaled families), a unified admissibility analysis in which one inequality on $G$ controls hyperbolicity, energy positivity, the potential bound along the kink, and the vacuum masses, and a numerical spectral study establishing the absence of negative modes across the admissible region.

Finally, the Yang–Mills connection has been reformulated in terms that are both honest and constructive. Exact SDYM reductions yield integrable equations, and the deformed model — like the $\phi^4$ model itself — is not integrable, so no exact reduction exists. The genuine bridge is the moduli-space dynamics of BPS monopoles and dyons descended from SDYM by dimensional reduction: any smooth one-dimensional moduli metric reduces, to second order, to the quadratic kinetic function studied here, and the Taub–NUT metric of well-separated dyon dynamics provides an explicit realization whose truncation lands at a computable point of the admissible plane. The completion of this embedding — the derivation of the sextic potential within a definite gauge-theory system — together with a proof of spectral stability, constitutes the natural continuation of this work.

# A. Direct verification of the inverse potential

We verify that $\phi_{\mathrm{k}}(x) = A\tanh(x/\Delta)$ solves the static field equation

$$G(\phi_{\mathrm{k}})\,\phi_{\mathrm{k}}{}'' \;+\; 1/2\;G'(\phi_{\mathrm{k}})\,\phi_{\mathrm{k}}{}'^2 \;=\; V'(\phi_{\mathrm{k}}) \tag{A.1}$$

for the potential in Eq. (13) with $V_0 = 0$.

Along the profile, the elementary identities

$$\phi_{\mathrm{k}}{}'^2 \;=\; \frac{A^2}{\Delta^2}\left(1-\frac{\phi_{\mathrm{k}}^2}{A^2}\right)^2, \qquad \phi_{\mathrm{k}}{}'' \;=\; -\frac{2\phi_{\mathrm{k}}}{\Delta^2}\left(1-\frac{\phi_{\mathrm{k}}^2}{A^2}\right) \tag{A.2}$$

hold, the second following from $\phi_{\mathrm{k}}{}'' = \phi_{\mathrm{k}}{}'\,(\mathrm{d}\phi_{\mathrm{k}}{}'/\mathrm{d}\phi_{\mathrm{k}})$. The left-hand side of Eq. (A.1) is therefore

$$-\frac{2\phi_{\mathrm{k}}}{\Delta^2}\,G(\phi_{\mathrm{k}})\left(1-\frac{\phi_{\mathrm{k}}^2}{A^2}\right) \;+\; \frac{A^2}{2\Delta^2}\,G'(\phi_{\mathrm{k}})\left(1-\frac{\phi_{\mathrm{k}}^2}{A^2}\right)^2. \tag{A.3}$$

On the other hand, differentiating (13),

$$V'(\phi) \;=\; \frac{A^2}{2\Delta^2}\left[G'(\phi)\left(1-\frac{\phi^2}{A^2}\right)^2 - \frac{4\phi}{A^2}\,G(\phi)\left(1-\frac{\phi^2}{A^2}\right)\right], \tag{A.4}$$

which coincides term by term with Eq. (A.3) evaluated at $\phi = \phi_{\mathrm{k}}$. Hence the profile satisfies the field equation exactly, for all $(\alpha, \beta)$ and all $(A, \Delta)$. The endpoints are regular: $\phi_{\mathrm{k}}{}' \to 0$ and $V - V_0 \to 0$ exponentially, so the kink approaches the vacua smoothly, and the local expansion about $\phi = \pm A$ reproduces the vacuum masses of Eq. (17).

An independent numerical verification of all closed-form results of this paper — the field equation, the quadrature, the canonical limits, the vacuum masses, the energy in Eq. (21), the fluctuation potential in Eq. (38), the zero-mode equation in Eq. (37), the asymptotics in Eq. (39), and both propositions — was carried out by computer-algebra substitution and by the finite-difference spectra of Section 6; all identities hold to machine precision.

# B. Exact positivity test for the quadratic kinetic function

For $G(\phi) = 1 + \alpha\phi + \beta\phi^2$ on the closed interval $[-A, A]$:

*Case* $\beta > 0$ *(convex).* The minimum over the interval is attained either at an endpoint or at the vertex $\phi_\star = -\alpha/2\beta$, if the vertex lies inside the interval. Hence $G > 0$ on $[-A, A]$ if and only if

$$1 + \alpha A + \beta A^2 > 0\,, \qquad 1 - \alpha A + \beta A^2 > 0\,, \qquad \text{and} \quad (|\alpha| \geq 2\beta A \ \text{ or } \ \alpha^2 \leq 4\beta)\,, \tag{B.1}$$

where the last clause states that either the vertex is outside the interval, or, if inside, its value $G(\phi_\star) = 1 - \alpha^2/4\beta$ is positive.

*Case* $\beta = 0$ *(linear).* The minimum is at an endpoint; $G > 0$ on $[-A, A]$ iff $1 \pm \alpha A > 0$.

*Case* $\beta < 0$ *(concave).* The minimum over a closed interval is attained at an endpoint; again $G > 0$ on $[-A, A]$ iff $1 \pm \alpha A + \beta A^2 > 0$.

In the dimensionless variables $\bar{a} = \alpha A$, $\bar{b} = \beta A^2$ these are the conditions in Eq. (32) used in Section 5. Note that the global condition in Eq. (33) ($\beta > 0$, $\alpha^2 \leq 4\beta$) implies the interval condition for every $A$, as it must.

# C. Derivation of the fluctuation potential

## C.1 Quadratic action

Expand $\phi = \phi_{\rm k} + \eta$ in the Lagrangian density. The spatial term gives, to second order in $\eta$ (with $G, G', G''$ evaluated at $\phi_{\rm k}$),

$$1/2\ G(\phi_{\rm k} + \eta)\,(\phi_{\rm k}{}' + \eta')^2 \ = \ 1/2\ G\,\phi_{\rm k}{}'^2 \ + \ G\,\phi_{\rm k}{}'\eta' \ + \ 1/2\ G' \\ + \ 1/2\ G\,\eta'^2 \ + \ G'\,\phi_{\rm k}{}'\,\eta\,\eta' \ + \ 1/ \tag{C.1}$$

and the potential contributes $V + \ V'\eta + \ 1/2\,V''\eta^2$. The first-order pieces vanish on the kink solution. The quadratic Lagrangian density is therefore

$$\mathcal{L}^{(2)} \ = \ 1/2\ \eta_t^2 \ - \ 1/2\ G(\phi_{\rm k})\,\eta_x^2 \ - \ G'(\phi_{\rm k})\,\phi_{\rm k}{}'\,\eta\,\eta_x \\ - \ 1/4\ G''(\phi_{\rm k})\,\phi_{\rm k}{}'^2\eta^2 \ - \ 1/2\ V''(\phi_{\rm k})\,\eta^2\,. \tag{C.2}$$

Integrating the cross term by parts, $-\int G'\phi_{\rm k}{}'\,\eta\,\eta_x = -\,1/2\int G'\phi_{\rm k}{}'\ \partial_x(\eta^2) = +\,1/2\int\ \partial_x(G'\phi_{\rm k}{}')\,\eta^2$, absorbs it into an effective mass, giving the quadratic action in Eq. (34) with

$$U \ = \ V''(\phi_{\rm k}) \ - \ 1/2\ G''(\phi_{\rm k})\,\phi_{\rm k}{}'^2 \ - \ 1/2\ \partial_x(G'(\phi_{\rm k})\phi_{\rm k}{}')\,, \tag{C.3}$$

which, on expanding the total derivative, is precisely the fluctuation potential in Eq. (35) of Section 6; it is not listed again here.

## C.2 Zero mode

Differentiating the static field equation in Eq. (A.1) with respect to $x$ gives

$$G\,\phi_{\rm k}{}''' \ + \ 2\,G'\,\phi_{\rm k}{}'\,\phi_{\rm k}{}'' \ + \ 1/2\ G''\,\phi_{\rm k}{}'^3 \ = \ V''(\phi_{\rm k})\,\phi_{\rm k}{}'\,, \tag{C.4}$$

where we used $\partial_x G = G'\phi_{\rm k}{}'$ and $\partial_x G' = G''\phi_{\rm k}{}'$. On the other hand, expanding Eqs. (36)–(35) on the argument $\phi_{\rm k}{}'$,

$$\mathcal{L}\,\phi_k' = -G\,\phi_k''' - 2\,G'\,\phi_k'\,\phi_k'' - 1/2\,G''\,\phi_k'^3 + V''(\phi_k)\,\phi_k', \tag{C.5}$$

which vanishes identically by Eq. (C.4). The translation zero mode in Eq. (37) therefore holds for arbitrary smooth kinetic functions, not only quadratic ones.

### C.3 Closed form of $U(x)$

With $t = \tanh(x/\Delta)$, $s = \mathrm{sech}(x/\Delta)$, $\phi_k = At$, and $G(\phi_k) = 1 + \alpha At + \beta A^2 t^2$, a direct computation gives

$$V''(\phi_k) = \frac{2}{\Delta^2}[\beta A^2/2\,s^4 - 2\alpha A\,t\,s^2 - 4\beta A^2 t^2 s^2 + (1 + \alpha At + \beta A^2 t^2)(2 - 3s^2)], \tag{C.6}$$

while the kinetic-induced terms are

$$-1/2\,G''\phi_k'^2 - G'\phi_k'' = -\frac{\beta A^2}{\Delta^2}s^4 + \frac{2}{\Delta^2}(\alpha A\,t\,s^2 + 2\beta A^2 t^2 s^2). \tag{C.7}$$

Adding Eqs. (C.6) and (C.7), the $s^4$ terms cancel and the remainder combines into Eq. (38). In the canonical limit $\alpha = \beta = 0$ only the Pöschl–Teller term survives, and as $s \to 0$ at $x \to \pm\infty$ one obtains $U(\pm\infty) = (4/\Delta^2)(1 \pm \alpha A + \beta A^2) = m_\pm^2$, the vacuum masses in Eq. (17). All steps were verified symbolically; the identities hold as exact polynomial identities in $t$.

## Statements

**Funding.** This research did not receive any specific grant from funding agencies in the public, commercial, or not-for-profit sectors.

**Conflict of interest.** The author declares no known competing financial interests or personal relationships that could have appeared to influence the work reported in this paper.

**Author contributions.** Ji Wang: conceptualization, methodology, formal analysis, investigation, software, writing — original draft, writing — review and editing.

**Data availability.** All data supporting the findings of this study are contained within the article. The symbolic and numerical verification procedures are described in Appendix A and can be reproduced with the parameters given in the text.